\documentstyle[prb,aps]{revtex}

\begin{document}

\twocolumn[\hsize\textwidth\columnwidth\hsize\csname 
@twocolumnfalse\endcsname

\title{Four-fold basal plane anisotropy of the nonlocal 
magnetization of YNi$_2$B$_2$C }
\author{L. Civale,$^1$  A.V. Silhanek,$^1$  J. R. Thompson,$^{2,3}$ K.J. Song,$^{3}$ C. V. Tomy,$^4$ and D. McK. Paul$^4$  }

\address{$^1$Comisi\'{o}n Nacional de Energ\'{\i}a At\'{o}mica-Centro At\'{o}mico 
Bariloche, 8400 Bariloche, Argentina}

\address{$^2$Oak Ridge National Laboratory, Oak Ridge, Tennessee 37831-6061 }

\address{$^3$Department of Physics, University of Tennessee, Knoxville, Tennessee 37996-1200}

\address{$^4$Department of Physics, University of Warwick, Coventry, CV4 7AL, UK}

\date{\today}

\maketitle 

\begin{abstract}
Studies of single crystal YNi$_2$B$_2$C have revealed a four-fold anisotropy of the equilibrium magnetization in the square crystallographic basal plane.  This $\pi /2$ periodicity occurs deep in the superconductive mixed state.  In this crystal symmetry, an ordinary superconductive mass anisotropy (as in usual London theory) allows only a constant, isotropic response.  In contrast, the experimental results are well described by generalized London theory incorporating non-local electrodynamics, as needed for this clean, intermediate-$\kappa$ superconductor.

\end{abstract}


\vskip1pc] \narrowtext

Borocarbide superconductors have received considerable recent attention, due
in part to the interaction between magnetism and superconductivity. A rich
superconducting phase diagram, including transitions between hexagonal, rhombohedral,
and square vortex lattices has been observed\cite{Yethiraj97a,Eskildsen97a,Paul98a,DeWilde97a}. 
The existence of vortex lattices with non-hexagonal symmetry has been attributed to nonlocality
effects on the superconducting electrodynamics\cite{DeWilde97a,Kogan97a}, which arise from the large electronic mean free path of these clean superconductors. Geometrically, a vortex directed along the tetragonal $\em{c}$-axis has current contours that are square-like \cite{Yethiraj98a}. It has been shown \cite{Song99a} in the non-magnetic borocarbide YNi$_2$B$_2$C, that the deviations from the standard (local) London magnetic field
dependence of the equilibrium magnetization $M_{eq} \propto$ ln$(H)$ can be quantitatively accounted for by introducing non-local electrodynamics into the London description \cite{Kogan96a}. Traditionally, it was widely thought that nonlocality effects should be significant only in materials with a Ginzburg-Landau parameter $\kappa$ $\sim$ 1; however, YNi$_2$B$_2$C has $\kappa$ $\approx$ 10 -- 15.

In the local London model of superconducting vortices, the material
anisotropy is introduced via a second rank mass tensor $m_{ij}$. In tetragonal materials such as YNi$_2$B$_2$C or LuNi$_2$B$_2$C, the masses in both principal directions in the square basal plane are the same, $m_a = m_b$, thus the superconducting properties are isotropic in the {\em a-b} plane. In contrast, non-local corrections are expected to introduce \cite{Miranovic99a} a four-fold anisotropy as a function of the magnetic field orientation within the {\em a-b} plane. A temperature dependent
in-plane anisotropy of the upper critical field $H_{c2}$ has been observed\cite{Metlushko97a} 
in the non-magnetic borocarbide LuNi$_2$B$_2$C and described 
within a Ginzburg-Landau framework incorporating non-local effects. However, a direct observation of the in-plane anisotropy deep into the superconducting
phase, where the non-local London model applies and unusual vortex lattices are observed,
has not been reported until now.

In this work we show that, in the superconducting mixed state of YNi$_2$B$_2$C,
the reversible magnetization oscillates with a $\pi/2$ periodicity when
the applied field is rotated within the 
{\em a-b} plane. The amplitude of the angular oscillation decreases with
field, passes through zero and then {\em reverses sign} at a
field well below $H_{c2}$. The results are in good quantitative agreement
with the non-local London description introduced by Kogan et al. \cite
{Kogan96a}

The 17 mg single crystal of YNi$_2$B$_2$C investigated in this study is the
same as that previously used by Song et al. \cite{Song99a} to explore the
magnetic response when the applied field {\bf H} is parallel to the {\em c}-axis. 
The critical temperature is $T_c = 14.5 K$, defined as the point at which the linearly varying magnetization $M(T)$ extrapolates to zero; this ignores a slight "tail" extending to $15.6 K$. The crystal is a slab of thickness $t \sim 0.5 mm$, whose shape and size in the basal plane are sketched in Fig. 1. It will be useful to approximate such shape by an ellipse of axis $L_x$ and $L_y$. X-ray diffraction shows that the crystallographic {\em c}-axis is normal to the slab, and that the two equivalent (110) axes of the tetragonal structure very approximately coincide with the axes of the ellipse. 

Measurements were performed in a
Quantum Design SQUID magnetometer with a 50 kOe magnet. Two sets of detection coils allows us to measure both the longitudinal (parallel to {\bf H}) and transverse (perpendicular to {\bf H}) components of the magnetization vector {\bf M}, but only the longitudinal component (denoted hereafter as simply the magnetization $M$) will be discussed in this work.
The crystal was mounted into a previously described\cite
{Silvia99a}, home-made rotating sample holder 
with rotation axis perpendicular to {\bf H}. The c-axis was aligned with the rotation
axis, so that {\bf H} could be rotated within the basal plane.

Figure 1 shows $M$ as a function of the angle $\varphi$ between {\bf H} and
the {\em a}-axis (see sketch in Fig. 1), at $T = 7 K$ and two values of $H$. The crystal was
initially cooled in zero field, $H$ was then applied and the sample
was subsequently rotated in steps of $\Delta \varphi
\approx 3.1 ^{\circ}$. Each data point was taken at fixed $\varphi$.

We first discuss the low field curve of fig. 1b. As $H = 30 Oe$ is well below
the lower critical field $H_{c1}$ for all $\varphi$, this curve represents the
total flux exclusion of the Meissner state. The oscillatory behavior with
periodicity $\pi$ (two-fold symmetry) originates from geometrical effects. 
Indeed, a field applied at any orientation within the basal plane can be decomposed in 
$H_x = H cos(\varphi+45^{\circ})$ and $H_y = H sin(\varphi+45^{\circ})$. If we approximate the crystal shape 
by the ellipse, the Meissner response associated with each component is $4\pi
M_i=-H_i/(1-\nu_i)$, where $i=x;y$ and $\nu_i=t/L_i$ are the demagnetizing factors, thus 
$4\pi M = -H [cos^2(\varphi+45^{\circ})/(1-\nu_x)+sin^2(\varphi+45^{\circ})/(1-\nu_y) ]$. The best fit to this expression gives $\nu_x \sim 1/4$ and $\nu_y \sim 1/5$. This corresponds to the ellipse of axes $L_x \sim 2.0 mm$ and $L_y \sim 2.5 mm$ shown in the sketch of fig. 1.

We now turn to the high field data of Fig. 1a. The applied field, $H = 45 kOe$, is well above $H_{c2}
\sim$ 35 kOe at this temperature (which is only weakly $\varphi$ dependent, see below), thus in this case $M(\varphi)$ = $M^{ns}(\varphi)$ is the normal state paramagnetic response. We again observe
an oscillatory behavior, but in this case the periodicity is $\pi/2$. By combining the information provided by the X-rays with the geometrical effects on the Meissner response, we conclude that the maximum normal state magnetization occurs at the crystallographic orientations 
$\langle 110 \rangle$ and $\langle {\overline1}10 \rangle$, while the minimum corresponds to 
$\langle 100 \rangle$ and $\langle 010 \rangle$. No hint of the geometry-originated two-fold symmetry
is observed. This is to be expected, as demagnetizing effects vanish in the limit $|M/H| \ll 1$.
Further analysis of $M^{ns}$ suggests that it arises from a low concentration, $\sim 0.001$ molar fraction, of rare earth ions, most likely from impurities in the yttrium starting metal.
As shown below, $M^{ns}$ is much smaller than the superconducting contribution except close to $H_{c2}$.

The above procedure cannot be used in the superconducting mixed state, due to the appearance of magnetic hysteresis arising from vortex pinning. The critical current density $J_c$ is very small in this crystal\cite{Song99a}. As a result, the magnetic hysteresis $({M^{\downarrow}}-{M^{\uparrow}}) \propto J_c$, where ${M^{\downarrow}}$ and 
${M^{\uparrow}}$ are respectively the magnetizations measured in the field-decreasing and field-increasing branches of an isothermal $M (H)$ loop, is small as compared to the 
equilibrium or reversible magnetization, $M_{eq} \approx ({M^{\downarrow}}+{M^{\uparrow}})/2$.
In spite of this, the residual hysteresis strongly affects the response obtained by rotating the crystal at fixed $T$ and $H$, by superimposing a periodicity $\pi$ (related to shape effects on the critical state magnetization) that almost completely hides the intrinsic $\pi /2$ periodicity of fundamental interest.

To solve this difficulty, we performed magnetization loops at $T = 7 K$
at a set of fixed angles and then calculated $M_{eq}(H)$ for each $\varphi$.
In all cases we extended the loops up to $H = 50 kOe$, thus we could repeat the measurement of $M^{ns}$ in the normal state and compare the data with those obtained by rotating at fixed 
$H$. Due to the absence of hysteresis, both determinations of 
$M^{ns}(H,\varphi)$ should coincide. This is indeed the case as seen in Fig.
1a, where the open circles represent the data at $H = 45 kOe$ obtained from
the $M(H)$ loops.

Figure 2 shows $M_{eq}$ (obtained from averaging $M^{\downarrow}$ and $M^{\uparrow}$) as a function of $\varphi$ for several $H$. All the data have the
same scale, but the curves at different $H$ have been vertically shifted to
accommodate the whole field range within the plot. For $H < 1.5 kOe$, the irreversibility becomes large enough to introduce a significant uncertainty in the determination of $M_{eq}$, consequently those data have been disregarded.
It is apparent that a four-fold symmetry exists in the whole field range of the measurements. To quantify the amplitude of the
oscillations, we fitted the curves by 
$M_{eq} (H,\varphi)$ = $\langle M_{eq} \rangle$ + $\delta M_{eq}(H) cos(4\varphi)$.

The values of $\delta M_{eq}(H)$ so obtained are plotted in Figure 3, while the values of $\langle M_{eq} \rangle (H)$ are shown in the inset. A remarkable fact, clearly visible in figs. 2 and 3, is that $\delta M_{eq}(H)$ crosses zero and {\em it reverses sign} at some intermediate field ($\sim$ 12 kOe) well within the superconducting mixed state. Another interesting observation is that the amplitude of the oscillations at $H \sim$ 1.5 - 2 kOe is as large as that at $H \sim 50 kOe$.

The above results show that a $\pi/2$ basal plane anisotropy exists both in
the normal and superconducting states. It is also clear from Fig. 3 that a
change in the behavior of $\delta M_{eq}(H)$ takes place at the
superconducting transition at $H_{c2} \sim 35 kOe$. This observation,
together with the sign reversal and the large amplitudes at low fields,
point to the existence of a second source of in-plane anisotropy, in addition to the
normal state one, that turns on in the superconducting phase.

Prior to analyzing the superconducting basal-plane anisotropy it is
necessary to subtract the normal state contribution, which persists within
the superconducting phase. To that end we performed rotations at fixed 
$H$, as those shown in Fig. 1a, at several $T$ and $H$ above 
$H_{c2}(T)$. We found a paramagnetic response that exhibits a four-fold symmetry, 
with the minimum at $\varphi =0$ in all cases, i.e., 
$M^{ns}(\varphi =45^{\circ})>M^{ns}(\varphi =0^{\circ})>0$ for all 
$T$ and $H$. We thus have a well defined set of data $\delta M^{ns} (H,T)$ 
which exhibits no sign reversal. The extrapolation is not obvious, however, as 
$\delta M^{ns}$ is not linear in $H$. Figure 4 shows all the $\delta M^{ns}$ 
data collected at various temperatures $5 K \le T \le 16 K$ and $H \le 50 kOe$, 
as a function of $H/T$. We found that, when plotted in this way, all
the data points collapse on a single curve.

The dashed line in Fig. 4 is a fit to the $\delta M^{ns} (H/T)$ data. The same fit, 
for the case $T = 7K$, is also shown as a dashed line in Fig. 3. 
We can now subtract that curve from the total $\delta M_{eq}$ shown in
Fig. 3, to isolate the superconducting contribution $\delta M_{eq}^{sc}$.
Note that, as $\delta M^{ns}$ is always positive and increases monotonically
with $H$, both the sign reversal and the non-monotonic behavior immediately
below $H_{c2}$ exhibited by $\delta M_{eq}$ must arise from the $\delta
M_{eq}^{sc}$ contribution.

We now show that the four-fold symmetry of $M_{eq}^{sc}$, as well as the field 
dependence of $\delta M_{eq}^{sc}$, can be well described using the non-local 
modifications to the London electrodynamics introduced by Kogan et al. \cite{Kogan96a}. According to that model, for $H_{c1} \ll H \ll H_{c2}$

\begin{equation}
M_{eq}^{sc}=-M_0 \left[ ln \left( \frac {H_0} {H} +1 \right) - \frac {H_0} {H_0 + H} +
\zeta \right]  \label{Kogan}
\end{equation}

Here $M_0 = \Phi_0/32\pi^2\lambda^2$, the new characteristic field $%
H_0=\Phi_0/4\pi^2\rho^2$ is related to the nonlocality radius $\rho$, and $%
\zeta = \eta_1-ln(H_0/\eta_2 H_{c2}+1)$, where $\eta_1$ and $\eta_2$ are
constants of order unity.

Song et al.\cite{Song99a} have shown that the magnetization of this same crystal is very
well described by Kogan's model, when $H \parallel {\it c}$-axis. A
fingerprint of the nonlocality effects is the deviation from the $M_{eq} \propto$
ln$(H)$ behavior predicted by the local London model. The curvature clearly 
visible in the inset of Fig. 3 is thus a strong indication that nonlocality 
also plays a major role when $H \perp {\it c}$-axis. It is worth mentioning that, 
although a quantitative analysis of the curve in the inset of Fig. 3 in terms of Eq. \ref{Kogan} would require removal of the normal state magnetization, its contribution is
small and would not significantly modify the curvature seen in the $M_{eq}$ vs. ln($H$) data.

We now apply Eq. \ref{Kogan} to the analysis of our data. In principle, the 
basal plane anisotropy could be ascribed
to the material parameters $M_0$, $H_0$ and $H_{c2}$. However, $M_0 \propto
\lambda^{-2}$ is isotropic within the {\em a-b} plane of a
tetragonal structure. On the other hand, four-fold variations of $H_{c2}$
within the basal plane have been observed in LuNi$_2$B$_2$C and attributed to 
nonlocality\cite{Metlushko97a}. We then assume that both $H_0$ and $H_{c2}$
have $\pi/2$ periodicity, $H_0(\varphi)=\langle H_0
\rangle + \delta H_0 cos(4\varphi)$ and $H_{c2}(\varphi)=\langle H_{c2}
\rangle + \delta H_{c2} cos(4\varphi)$. To first order in $\delta H_0$ and
$\delta H_{c2}$ we obtain

\begin{equation}
\delta M^{sc}_{eq}=-M_0 \left[ \left( \frac {1} {\left( 1+ \frac {H}{\langle
H_0 \rangle } \right)^2} -\alpha \right)\epsilon_1 + \alpha \epsilon_2 %
\right]  \label{oscillation}
\end{equation}
where $\; \; \; \; \alpha = \frac {1} {1 + \eta_2 \frac {\langle H_{c2} \rangle } { \langle H_0
\rangle } }; \; \; \; \; \; \;    
\epsilon_1 = \frac {\delta H_0} {\langle H_0 \rangle}; \; \; \; \; \; \; 
\epsilon_2 = \frac {\delta H_{c2}} {\langle H_{c2} \rangle}$
\\

Experimentally we have determined $\langle H_{c2} \rangle$ = 35 kOe and $\delta H_{c2}$ 
$\sim$ 0.4 kOe (at $T=7K$), so we can fix $\epsilon_2$ = 0.01. We could also attempt to determine $M_0$ and $\langle H_0 \rangle$ by fitting our $M_{eq}$ data with Eq. 
\ref{Kogan}. However, this is a difficult task that requires\cite{Song99a} the measurement 
of a large set of temperatures to check the consistency of the results. Instead, we decided 
to use the results of Song et al.\cite{Song99a}, (for $H \parallel {\em c}$-axis) as good estimates. 
For $T$ = 7 K, we take $\langle H_0 \rangle$ = 56 kOe 
and $M_0$ = 5.2 G. (Here we scaled down $M_0 \propto 1/ \lambda_a \lambda_c$ 
by the experimental mass anisotropy, $\gamma \approx 1.15$, between
the {\em c}-axis and the {\em ab}-plane, which is close to the value $\gamma \sim 1.1$ obtained from band structure calculations\cite{Singh96a}.)
In any case, small variations in any of these parameters will not
significantly affect the rest of the analysis. If we also assume $\eta_2 \sim 1$, we obtain $\alpha \sim 2/3$. With these fixed parameters, we fit our $\delta M^{sc}_{eq}(H)$ data with Eq. \ref{oscillation}, with the {\emph single} free parameter $\epsilon_1$. 
We obtain $\epsilon_1$ = 0.14. The fitted curve (for $\delta M^{sc}_{eq}+\delta M^{ns}$) is shown as a solid line in Fig. 3.

The very good coincidence between our data and the model is
remarkable. With a single fitting parameter $\epsilon_1$,
which is field independent, we have been able to account
for the nontrivial $H$ dependence of $\delta M^{sc}_{eq}$, including the
sign reversal at intermediate fields. Of course, the fit deviates from the data close to $H_{c2}$, where the London model fails.  These experimental results show that nonlocality effects have a profound effect on these clean, intermediate-$\kappa$ superconductors and they underscore the remarkable utility of the generalized London theory.

In summary, we have demonstrated a four-fold anisotropy in the square basal plane of clean single crystal YNi$_2$B$_2$C. This superconducting response is inconsistent with conventional local London theory, but it is well explained by a generalized London model incorporating non-local electrodynamics, with parameters based largely on complementary experiments.  These observed effects of nonlocality persist deep into the superconducting state, where complex, evolving vortex lattices occur.

[Note added:  while preparing this manuscript, we learned that P. C. Canfield et al. at the Ames  Lab. have observed similar oscillations of the basal plane magnetization of LuNi$_2$B$_2$C.]  

We are pleased to acknowledge useful discussions with V.G. Kogan. A.S. is member of CONICET.  Collaboration between UTK and CAB was supported in part by a UTK Faculty Research Fund.  Research at the ORNL is supported by the Div. of Materials Sciences, U.S. Dept. of Energy under contract number DE-AC05-96OR22464 with Lockheed Martin Energy Research Corp.

\noindent
\begin{figure}
\caption[]{Equilibrium magnetization $M_{eq}$ at $T = 7 K$, as a function 
of the angle $\varphi$ between the applied field {\bf H} (contained 
in the {\em a-b} plane) and the {\em a}-axis, for (a) $H = 45 kOe$ and (b) $H = 30 Oe$.
The inset shows a sketch of the crystal shape and crystallographic orientation in the basal plane, superimposed to an ellipse with the same demagnetizing factors.}
\label{Fig. 1}
\end{figure}

\noindent
\begin{figure}
\caption[]{Reversible magnetization $M_{eq}$ at $T = 7 K$,
as a function of the angle $\varphi$ between the applied field {\bf H} (contained 
in the {\em a-b} plane) and the {\em a}-axis. The fields (in kOe) are indicated next to each curve. The scale is the same for all the curves, but data at different $H$ have been vertically shifted.}
\label{Fig. 2}
\end{figure}

\noindent
\begin{figure}
\caption[]{Amplitude $\delta M_{eq}$ of the four-fold oscillations of the basal plane magnetization, as a function of the applied field $H$. The dashed line is the normal state contribution $\delta M^{ns}$ obtained from the fit shown in Fig. 4. The solid line is the fit to $\delta M^{sc}_{eq} + \delta M^{ns}$ using Eq. (2) with $\epsilon_1$ as the only fitting parameter. Inset: average in-plane magnetization $\langle M_{eq} \rangle$ as a function of applied field (see text).} 
\label{Fig. 3}
\end{figure}

\noindent
\begin{figure}
\caption[]{
Amplitude $\delta M^{ns}$ of the four-fold oscillations of the basal plane magnetization 
in the normal state, as a function of $H/T$. The dashed line is a polynomial fit.} 
\label{Fig. 4}
\end{figure}


\begin{references}
\bibitem{Yethiraj97a}  M. Yethiraj, D. McK. Paul, C. V. Tomy, and E. M.
Forgan, Phys. Rev. Lett. {\bf 78}, 4849 (1997).

\bibitem{Eskildsen97a}  M. R. Eskildsen, P. L. Gammel, B. P. Barber, A. P.
Ramirez, D. J., Bishop, N. H. Andersen, K. Mortensen, C. A. Bolle, C. M.
Lieber, and P. C. Canfield, Phys. Rev. Lett. {\bf 79},
487 (1997).

\bibitem{Paul98a}  D. McK. Paul, C. V. Tomy, C. M. Aegerter, R. Cubitt, S.
H. Lloyd, E. M. Forgan, S. L. Lee, and M. Yethiraj, Phys. Rev. Lett. {\bf 80}, 1517 (1998).

\bibitem{DeWilde97a}  Y. De Wilde, M. Iavarone, U. Welp, V. Metlushko, A. E.
Koshelev, I. Aronson, and G. W. Crabtree, Phys. Rev. Lett. {\bf 78}, 4273 (1997).

\bibitem{Yethiraj98a}  M. Yethiraj, D.McK Paul, C. V. Tomy, and J. R. Thompson, Phys. Rev. B 58, R14767 (1998).

\bibitem{Kogan97a}  V. G. Kogan, M. Bullock, B. Harmon, P. Miranovic, Lj.
Dobrosavljevic-Grujic, P. L. Gammel, and D. J. Bishop, 
Phys. Rev. B {\bf 55}, R8693 (1997).

\bibitem{Song99a}  K. J. Song, J. R. Thompson, M. Yethiraj, D. K. Christen,
C. V. Tomy, and D. McK. Paul, Phys. Rev. B {\bf 59}, R6620
(1999).

\bibitem{Kogan96a}  V. G. Kogan, A. Gurevich, J. H. Cho, D. C. Johnston,
Ming Xu, J. R. Thompson, and A. Martynovich, Phys. Rev. B 
{\bf 54}, 12386 (1996).

\bibitem{Miranovic99a}  V. G. Kogan, P. Miranovic, and D. McK. Paul, in \emph{The Superconducting State in Magnetic Fields:  Special Topics and Trends}, ed. by C. A. Sa de Melo, Directions in Condensed Matter Physics vol 13 (World Scientific, Singapore, 1998)

\bibitem{Metlushko97a}  V. Metlushko, U. Welp, A. Koshelev, I. Aranson, G.
W. Crabtree, and P. C. Canfield, 
Phys. Rev. Lett. {\bf 79}, 1738 (1997).

\bibitem{Silvia99a}  S. Candia and L. Civale, Supercond. Sci. Technol. {\bf 12}, 192 (1999); A. Silhanek, L. Civale, S. Candia, G. Pasquini and G. Nieva, Phys. Rev. B {\bf 59}, 13620 (1999).

\bibitem{Singh96a}  D. J. Singh, Sol. St. Comm. {\bf 98}, 899 (1996).

\end{references}
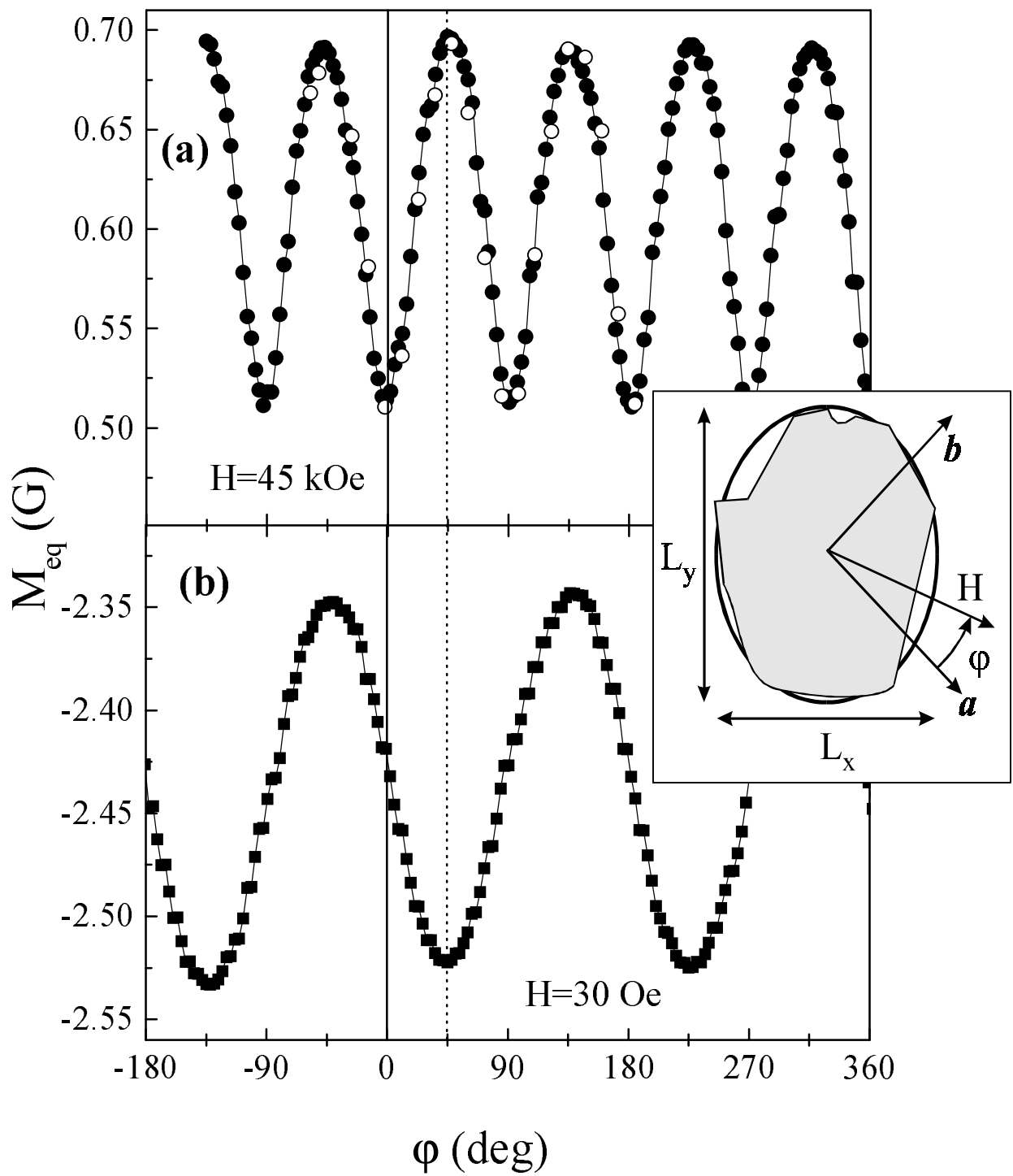
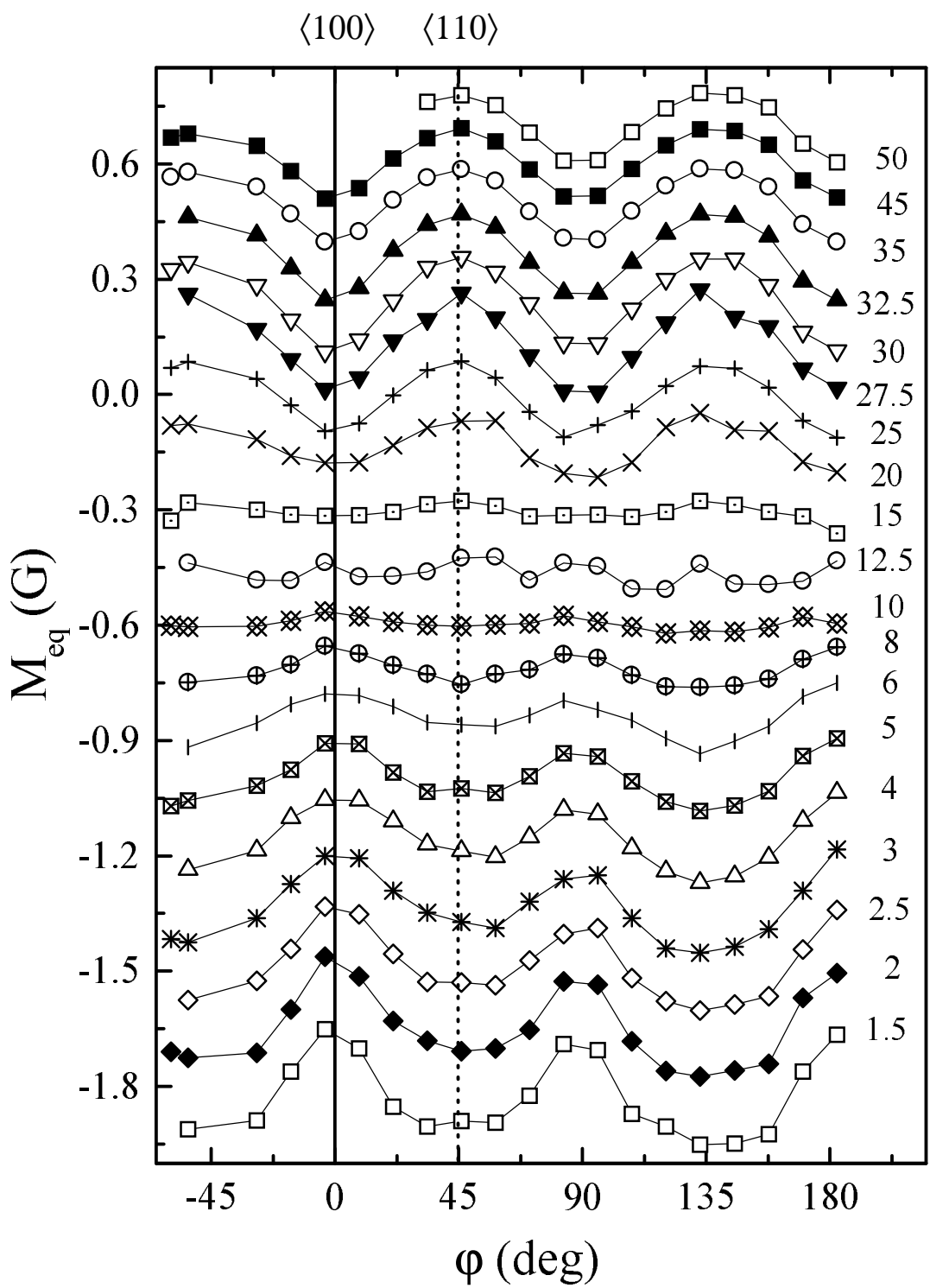
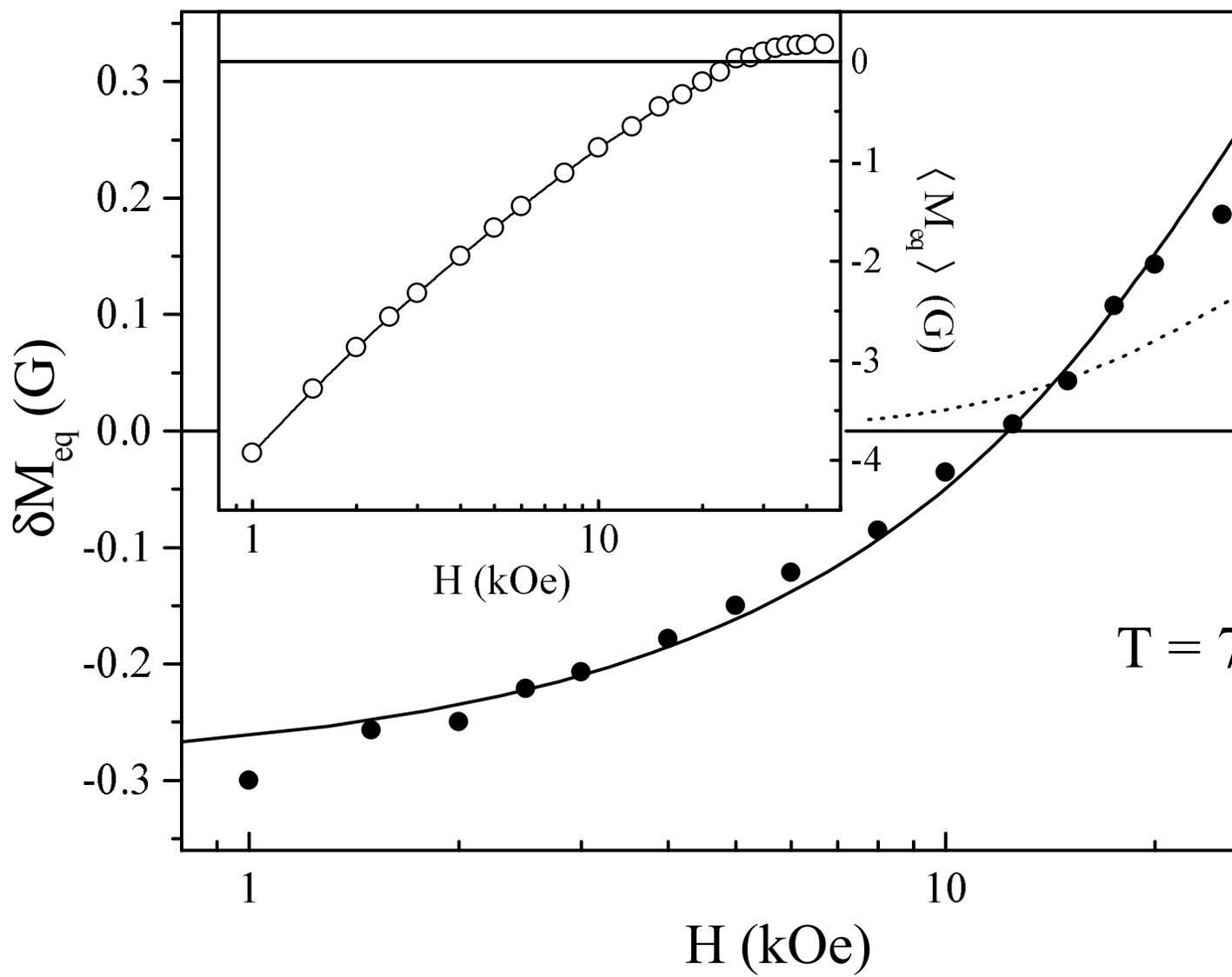
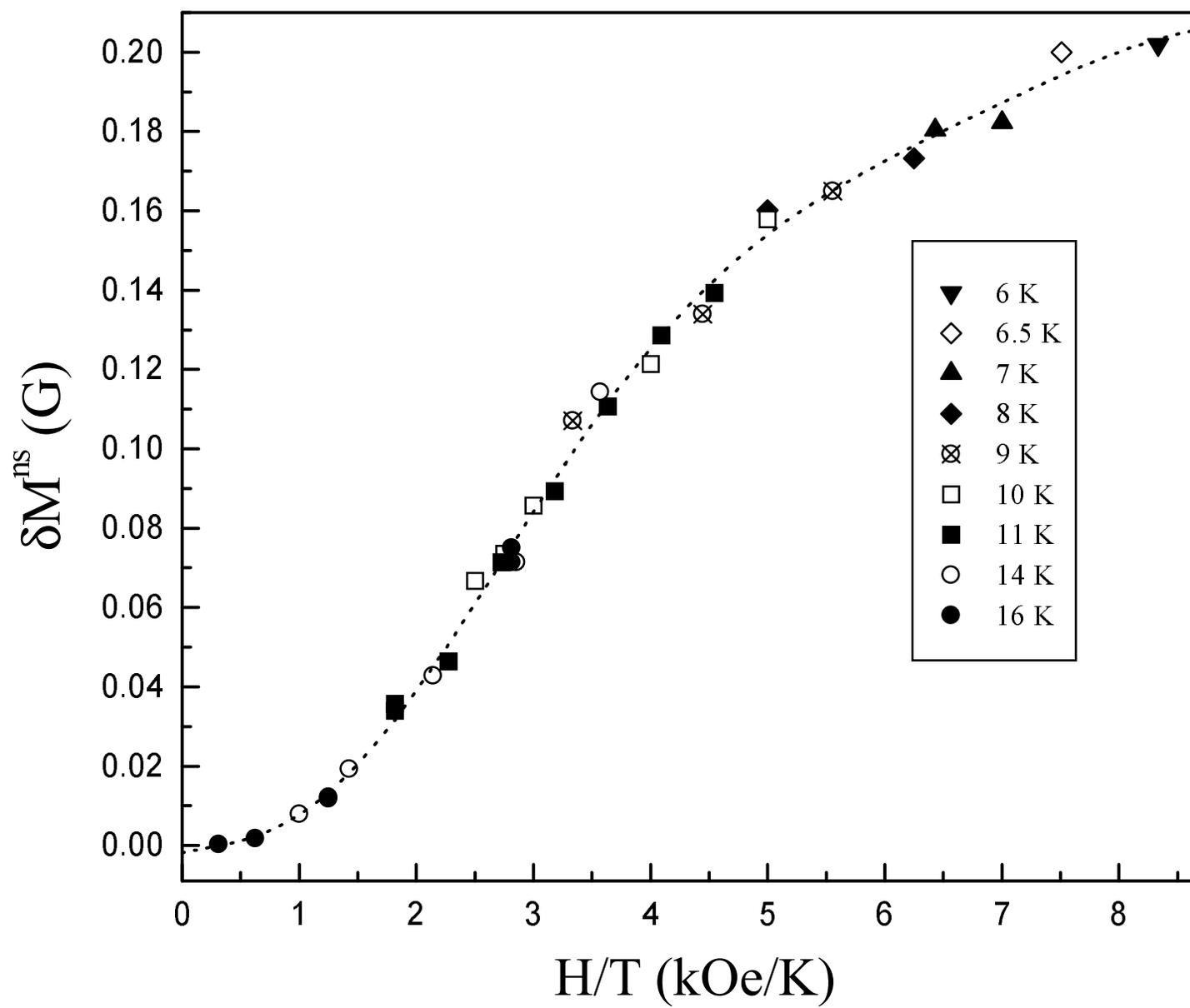
\end{document}